\documentclass{article}
\usepackage{spconf,amsmath,graphicx}
\usepackage{booktabs}
\usepackage{amssymb}
\usepackage{bm}
\usepackage{multirow}
\usepackage{arydshln}
\usepackage{cite}
\usepackage{setspace}
\usepackage{array}
\usepackage[hyphens]{url}
\usepackage{cancel}
\usepackage{algorithm,algorithmicx,algpseudocode}
\usepackage{xcolor,soul}
\sethlcolor{lightgray}

\definecolor{red}{rgb}{1,0,0}
\newcommand{\red}[1]{{\color{red} #1}}
\definecolor{blue}{rgb}{0,0,1}
\newcommand{\blue}[1]{{\color{blue} #1}}

\makeatletter
\def\adl@drawiv#1#2#3{%
        \hskip.5\tabcolsep
        \xleaders#3{#2.5\@tempdimb #1{1}#2.5\@tempdimb}%
                #2\z@ plus1fil minus1fil\relax
        \hskip.5\tabcolsep}
\newcommand{\cdashlinelr}[1]{%
  \noalign{\vskip\aboverulesep
           \global\let\@dashdrawstore\adl@draw
           \global\let\adl@draw\adl@drawiv}
  \cdashline{#1}
  \noalign{\global\let\adl@draw\@dashdrawstore
           \vskip\belowrulesep}}
\makeatother

\newcolumntype{x}[1]{>{\centering\arraybackslash\hspace{0pt}}p{#1}}

\usepackage{tikz}
\usepackage{pgfplots}
\pgfplotsset{width=7.5cm,compat=1.12}
\usepgfplotslibrary{fillbetween}

\renewcommand{\paragraph}[1]{\noindent\textbf{#1}\hspace{0.5em}}

\title{BECTRA: Transducer-based End-to-End ASR with BERT-Enhanced Encoder}
\name{
    Yosuke Higuchi$^{1}$,
    Tetsuji Ogawa$^{1}$,
    Tetsunori Kobayashi$^{1}$,
    Shinji Watanabe$^{2}$
}

\address{
    $^1$Waseda University, Japan\ \ $^2$Carnegie Mellon University, USA
 }

\begin{document}
\ninept
\maketitle
\setlength{\abovedisplayskip}{4pt}
\setlength{\belowdisplayskip}{4pt}
\setlength{\textfloatsep}{0.4cm} %
\begin{abstract}
We present \textbf{BE}RT-\textbf{C}TC-\textbf{Tra}nsducer (BECTRA),
a novel end-to-end automatic speech recognition (E2E-ASR) model
formulated by the transducer with a BERT-enhanced encoder.
Integrating a large-scale pre-trained language model (LM) into E2E-ASR has been actively studied,
aiming to utilize versatile linguistic knowledge for generating accurate text.
One crucial factor that makes this integration challenging lies in the vocabulary mismatch;
the vocabulary constructed for a pre-trained LM is generally too large for E2E-ASR training and
is likely to have a mismatch against a target ASR domain.
To overcome such an issue, we propose BECTRA,
an extended version of our previous BERT-CTC,
that realizes BERT-based E2E-ASR using a vocabulary of interest.
BECTRA is a transducer-based model,
which adopts BERT-CTC for its encoder and
trains an ASR-specific decoder using a vocabulary suitable for a target task.
With the combination of the transducer and BERT-CTC,
we also propose a novel inference algorithm
for taking advantage of both autoregressive and non-autoregressive decoding.
Experimental results on several ASR tasks,
varying in amounts of data, speaking styles, and languages,
demonstrate that BECTRA outperforms BERT-CTC by
effectively dealing with the vocabulary mismatch while exploiting BERT knowledge.
\end{abstract}

\vspace{-0.1cm}
\section{Introduction}
\vspace{-0.1cm}
In the field of natural language processing (NLP),
language model (LM) pre-training has achieved remarkable success and become a dominant paradigm.
With well-designed self-supervised objectives and a dramatic increase in model capacity,
large-scale LMs~\cite{devlin2019bert,brown2020language} are pre-trained on a vast amount of text-only data to acquire versatile linguistic knowledge~\cite{tenney2019bert}.
Such pre-trained LMs (PLMs) provide rich representations for boosting the performance of downstream NLP tasks
while alleviating the heavy requirement of supervised data.

Inspired by the great success in NLP,
there has been an increasing interest in adopting a PLM for
end-to-end automatic speech recognition (E2E-ASR).
Several works have introduced PLMs to E2E-ASR via
N-best hypothesis rescoring~\cite{shin2019effective,salazar2020masked,chiu2021innovative,futami2021asr,udagawa2022effect} and
knowledge distillation~\cite{futami2020distilling,bai2021fast,kubo2022knowledge,lu2022context}.
More recently,
others have attempted to fine-tune a PLM directly for E2E-ASR~\cite{huang2021speech,yi2021efficiently,zheng2021wav,deng2021improving,yu2022non}.
This enables a model to exploit the LM's powerful representations during training and inference
while requiring a complex mechanism for bridging the speech-text gap and
careful adjustment for stabilizing the fine-tuning process.

BERT-CTC is another possibility we have explored for incorporating a pre-trained masked LM, i.e., BERT, into the formulation of connectionist temporal classification (CTC)~\cite{higuchi2022bert}.
BERT-CTC efficiently uses BERT \textit{without fine-tuning}
to explicitly condition CTC on context-aware linguistic information,
alleviating the conditional independence in the output dependency.
While BERT-CTC has shown promising results,
its performance is still limited due to a discrepancy in its output vocabulary.
BERT-CTC predicts a sequence via a masked LM-based iterative refinement algorithm~\cite{ghazvininejad2019mask} (e.g., Table~\ref{tb:decoding_example}) with repeated updates on BERT embeddings.
This BERT-based refinement forces the model to be trained on the BERT vocabulary,
which is often too large for ASR training and is likely to have a mismatch against a target ASR domain.
In fact, as shown in~\cite{higuchi2022bert},
the BERT-CTC performance becomes less significant
when compared with standard CTC and transducer-based models
trained on a vocabulary constructed from ASR training text.
A similar issue has been reported in~\cite{deng2022improving},
which has mentioned that alphabetic languages (e.g., English) are prone to the vocabulary mismatch due to the difference in subword units.

In this work, we propose \textbf{BE}RT-\textbf{C}TC-\textbf{Tra}nsducer (BECTRA),
an extension of BERT-CTC that is capable of handling the vocabulary mismatch and exploiting BERT knowledge for improving ASR performance.
BECTRA formulates E2E-ASR based on the transducer~\cite{graves2012sequence},
which comprises an encoder enhanced by BERT-CTC and
a decoder (i.e., prediction/joint networks) trained with an ASR-specific vocabulary.
Such a separate decoder enables a model to learn text generation more accurately in a suitable output unit,
still benefiting from the BERT-conditioned framework.
Moreover, BECTRA models E2E-ASR in a more accurate formulation than BERT-CTC,
thanks to the transducer's autoregressive characteristic relying less on conditional independence assumptions.
With the combination of the transducer and BERT-CTC,
we also propose a novel inference algorithm for BECTRA,
which takes advantage of both autoregressive and non-autoregressive decoding algorithms.

\vspace{-0.2cm}
\section{Background: BERT-CTC for end-to-end ASR}
\vspace{-0.15cm}
Let $O\!=\!(\bm{\mathrm{o}}_t\!\in\!\mathbb{R}^F| t\!=\!1,\cdots,T)$ be an input sequence of length $T$, and
$W^{\mathsf{b}} \!=\!( w^{\mathsf{b}}_n\!\in\!\mathcal{V}^{\mathsf{b}} | n\!=\!1,\cdots,N )$ be the corresponding output sequence of length $N$,
where $\bm{\mathrm{o}}_t$ is an $F$-dimensional acoustic feature at frame $t$, 
$w^{\mathsf{b}}_n$ is an output token at position $n$, and $\mathcal{V}^{\mathsf{b}}$ is a vocabulary of BERT~\cite{devlin2019bert} trained on a target language.
Note that the superscript (${\mathsf{b}}$) indicates the use of the BERT vocabulary.
The objective of E2E-ASR is to formulate the direct mapping from $O$ to $W^{\mathsf{b}}$ by
modeling a posterior distribution $p(W^{\mathsf{b}}|O)$ using a single deep neural network.

BERT-CTC~\cite{higuchi2022bert} formulates $p(W^{\mathsf{b}}|O)$ by introducing a masked sequence
$\tilde{W}^{\mathsf{b}}\!=\!(\tilde{w}^{\mathsf{b}}_n\!\in\!\mathcal{V}^{\mathsf{b}} \cup \{\varnothing\} | n\!=\!1,\cdots,N)$,
which is obtained by replacing some tokens in an output sequence $W^{\mathsf{b}}$ with a special mask token $\varnothing$ (see Table~\ref{tb:decoding_example} for example).
Considering all possible masking patterns compatible with $W^{\mathsf{b}}$,
$p(W^{\mathsf{b}}|O)$ is factorized as
\begin{equation}
    p(W^{\mathsf{b}}|O) = \sum_{\tilde{W}^{\mathsf{b}}} p(W^{\mathsf{b}}|\tilde{W}^{\mathsf{b}},O) p(\tilde{W}^{\mathsf{b}}|O). \label{eq:p_bc_W_O}
\end{equation}
Similar to CTC~\cite{graves2006connectionist},
$p(W^{\mathsf{b}}|\tilde{W}^{\mathsf{b}},O)$ in Eq.~\eqref{eq:p_bc_W_O} is marginalized over all possible alignment paths between $O$ and $W^{\mathsf{b}}$ as
\begin{equation}
    p(W^{\mathsf{b}}|\tilde{W}^{\mathsf{b}},O) \approx 
    \sum_{A^{\mathsf{b}}\in\mathcal{B}^{\text{--}1}(W^{\mathsf{b}})} p(A^{\mathsf{b}}|W^{\mathsf{b}},O) p(W^{\mathsf{b}}|\tilde{W}^{\mathsf{b}}),
    \label{eq:p_bc_W_tW_O}
\end{equation}
where $A^{\mathsf{b}}\!=\!(a^{\mathsf{b}}_t\!\in\!\mathcal{V}^{\mathsf{b}} \cup \{\epsilon\} | t\!=\!1,\cdots,T)$ is a frame-level token sequence obtained by augmenting $W^{\mathsf{b}}$ with a blank symbol $\epsilon$, and
$\mathcal{B}:A^{\mathsf{b}}\!\mapsto\!W^{\mathsf{b}}$ is the collapsing function~\cite{graves2006connectionist} that removes token repetitions and blank symbols from $A^{\mathsf{b}}$.
To obtain Eq.~\eqref{eq:p_bc_W_tW_O},
BERT-CTC assumes reasonable conditional independence of $\tilde{W}^{\mathsf{b}}$ and $O$.
The alignment probability $p(A^{\mathsf{b}}|W^{\mathsf{b}},O)$ is further factorized using the probabilistic chain rule as
\begin{equation}
    \text{Eq.~\eqref{eq:p_bc_W_tW_O}}
    \approx \sum_{A^{\mathsf{b}}} \prod_{t=1}^{T} p(a^{\mathsf{b}}_t|\cancel{a^{\mathsf{b}}_1,\cdots,a^{\mathsf{b}}_{t-1}},W^{\mathsf{b}},O) p(W^{\mathsf{b}}|\tilde{W}^{\mathsf{b}}),
    \label{eq:p_bc_W_tW_O_2}
\end{equation}
which makes the same conditional independence assumption identical to CTC.
Assuming  $p(W^{\mathsf{b}}|\tilde{W}^{\mathsf{b}})$ as a strong prior probability modeled by a pre-trained masked LM,
BERT-CTC models the product of $p(a^{\mathsf{b}}_t|W^{\mathsf{b}},O)$ and $p(W^{\mathsf{b}}|\tilde{W}^{\mathsf{b}})$ in Eq.~\eqref{eq:p_bc_W_tW_O_2} as
\begin{equation}
    \text{Eq.~\eqref{eq:p_bc_W_tW_O_2}} \triangleq \sum_{A^{\mathsf{b}}} \prod_{t} p(a^{\mathsf{b}}_t|\text{BERT}(\tilde{W}^{\mathsf{b}}),O), \label{eq:p_bc_W_tW_O_3}
\end{equation}
where $\text{BERT}(\cdot)$ indicates contextual embedding obtained from the output of BERT~\cite{devlin2019bert}.
Different from standard CTC formulation, in Eq.~\eqref{eq:p_bc_W_tW_O_3},
BERT-CTC conditions the token emission probability on BERT's contextual linguistic information.
This enables a model to not only utilize BERT knowledge for E2E-ASR
but also explicitly relax the conditional independence assumption made in CTC.
For more detailed derivation from Eq.~\eqref{eq:p_bc_W_tW_O} to Eq.~\eqref{eq:p_bc_W_tW_O_3},
see Sec.\ 3 of~\cite{higuchi2022bert}.

\begin{figure}[t]
    \centering
    \includegraphics[width=0.82\columnwidth]{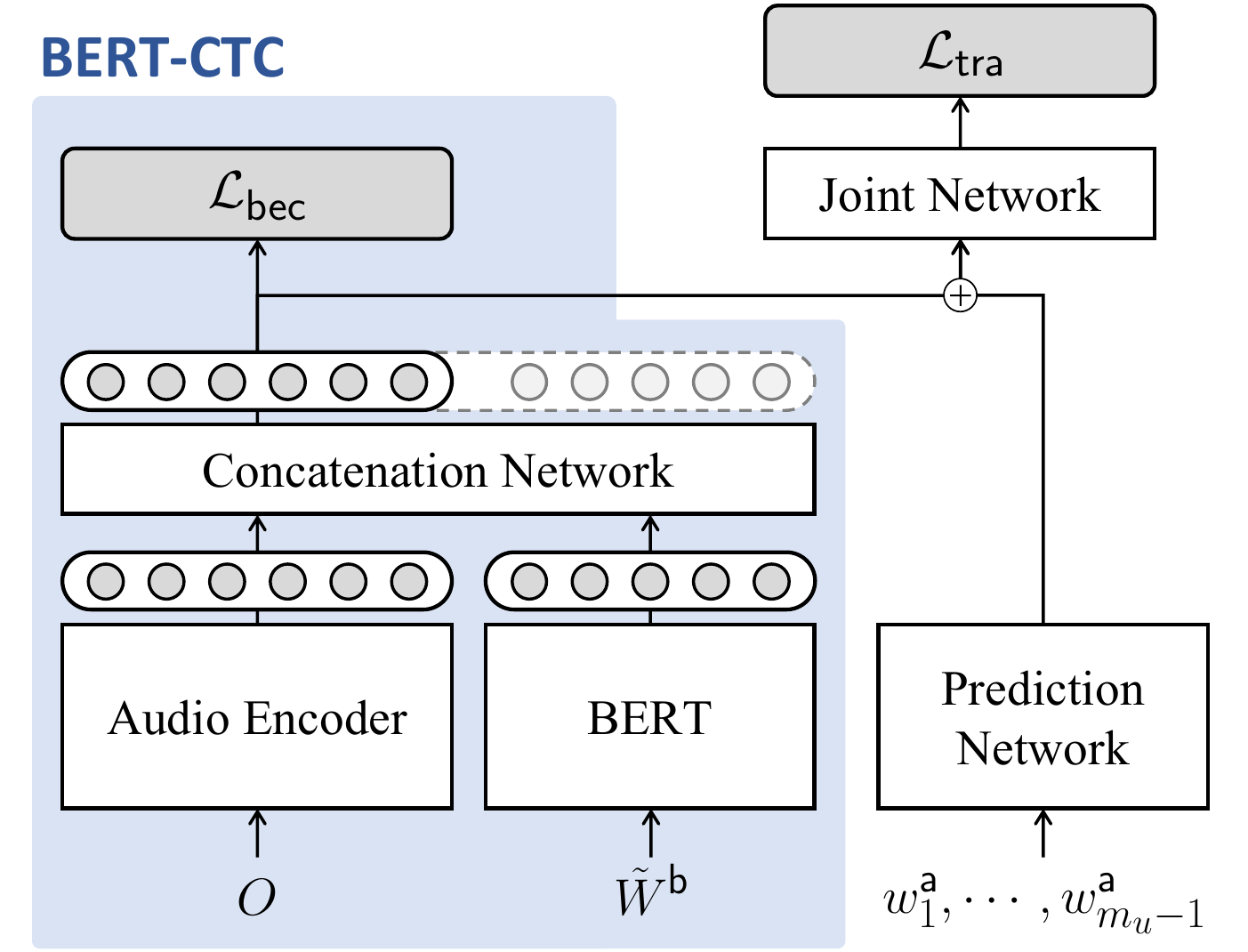}
    \vspace{-0.2cm}
    \caption{Overview of BECTRA architecture. BECTRA adopts BERT-CTC as its encoder for transducer-based E2E-ASR training.}
    \label{fig:bectra}
\end{figure}

The blue region in Fig.~\ref{fig:bectra} depicts the BERT-CTC architecture, consisting of an audio encoder, BERT, and a concatenation network.
The token emission probability in Eq.~\eqref{eq:p_bc_W_tW_O_3} is computed as
\begin{align}
    &\hspace{-0.06cm}p(a^{\mathsf{b}}_t|\text{BERT}(\tilde{W}^{\mathsf{b}}),O) = \sigma(\text{ConcatNet}_t(E, H)) \in [0, 1]^{|\mathcal{V}^{\mathsf{b}}|+1}, \hspace{-0.02cm} \label{eq:p_bc_at} \\
    &\hspace{-0.06cm}E = \text{AudioEnc}(O) \in \mathbb{R}^{T \times D},\ \ H = \text{BERT}(\tilde{W}^{\mathsf{b}}) \in \mathbb{R}^{N \times D},
\end{align}
where $D$ is the dimension of hidden vectors $\bm{\mathrm{e}}_t\!\in\!E$ and $\bm{\mathrm{h}}_n\!\in\!H$ in each encoded sequence, and
$\sigma(\cdot)$ is a softmax layer.
In Eq.~\eqref{eq:p_bc_at},
$\text{ConcatNet}_t(\cdot)$ indicates the $t$-th output of the concatenation network,
which adopts the self-attention mechanism~\cite{vaswani2017attention} for learning inner/inter dependencies within/between the concatenated $E$ and $H$.

\paragraph{Training}
The ojective function of BERT-CTC is defined by the negative log-likelihood of Eq.~\eqref{eq:p_bc_W_O} expanded with Eq.~\eqref{eq:p_bc_W_tW_O_3}:
\begin{align}
    \mathcal{L}_{\mathsf{bec}} &= -\log \sum_{\tilde{W}^{\mathsf{b}}} \sum_{A^{\mathsf{b}}} \prod_{t} p(a^{\mathsf{b}}_t|\text{BERT}(\tilde{W}^{\mathsf{b}}),O) p(\tilde{W}^{\mathsf{b}}|O) \nonumber \\
    &\le -\mathbb{E}_{\tilde{W}^{\mathsf{b}} \sim \mathcal{M}(W^{\mathsf{b}})} \bigg[ \log \sum_{A^{\mathsf{b}}} \prod_{t} p(a^{\mathsf{b}}_t|\text{BERT}(\tilde{W}^{\mathsf{b}}),O) \bigg], \label{eq:L_bec}
\end{align}
where the intractable marginalization over $\tilde{W}^{\mathsf{b}}$ is rewritten
under expectation with respect to a sampling distribution $\mathcal{M}(W^{\mathsf{b}})$.
Following~\cite{ghazvininejad2019mask},
the sampling process is performed by
first sampling the number of tokens from a uniform distribution as $M\!\sim\!\mathcal{U}(1,N)$ and then randomly masking $M$ tokens in a ground-truth sequence $W^{\mathsf{b}}$.
The summation over $A^{\mathsf{b}}$ is efficiently computed as in CTC~\cite{graves2006connectionist}.

\paragraph{Inference}
Algorithm~\ref{algo:bertctc_inference} represents a pseudocode for BERT-CTC inference,
which is inspired by the mask-predict algorithm~\cite{ghazvininejad2019mask} combined with CTC inference~\cite{chan2020imputer,higuchi2020mask,higuchi2021improved}.
The algorithm gradually generates an output sequence $\hat{W}^{\mathsf{b}}$ by
iterating the following procedure $K$ times.
At each iteration $k\!\in\!\{1,\cdots,K\}$,
a masked sequence $\hat{W}^{\mathsf{b}}$ is fed into BERT to obtain contextual embeddings $H$ (line 4).
Given $H$ and an encoder output $E$,
$\hat{W}^{\mathsf{b}}$ is updated with a newly predicted sequence,
which is obtained via the best path decoding~\cite{graves2006connectionist} using the framewise token probability from Eq.~\eqref{eq:p_bc_at} (line 5).
Then, $M$ tokens having the lowest probability scores are replaced with the mask token $\varnothing$ (lines 6 and 7),
where $M$ is calculated from a linear decay function $\lfloor |\hat{W}^{\mathsf{b}}| \cdot \frac{K - k}{K} \rfloor$ similar to~\cite{ghazvininejad2019mask}.

\vspace{-0.1cm}
\section{BECTRA}
\vspace{-0.2cm}

\begin{table*}[t]
    \centering
    \begin{tabular}{c@{\hspace{0.15cm}}ll}
        \toprule
        \multirow{4}{*}[-1pt]{\begin{rotatebox}{90}{$\xleftarrow{\hspace*{1.1cm}}$}\end{rotatebox}}&
        BERT-CTC ($k\!=\!1$) & ... \hl{thou} \hl{again} \hl{meet} \hl{any} \hl{one} after \hl{thee} \hl{hour} \hl{reciting} \hl{artht} \hl{of} \hl{poetryry} \hl{whether} \hl{he} \hl{be} \hl{near'or} \hl{far} \hl{it} \hl{will} \hl{be} \hl{a} ... \\
        & BERT-CTC ($k\!=\!5$) & ... thou \hl{again} \hl{meet} \hl{any} \hl{one} after \hl{this} \hl{hour} \hl{reciting} \hl{artht} of \hl{poetryry} whether \hl{he} \hl{be} \hl{near} or \hl{far} it \hl{will} be \hl{i} ... \\
        & BERT-CTC ($k\!=\!10$) & ... thou again meet any one after \blue{this} hour reciting \blue{aught} of \red{poetryry} whether he be \blue{near or} far it will be \blue{i} ... \\
        & BECTRA ($B\!=\!5$) & ... thou again meet any one after this hour reciting aught of \blue{poetry} whether he be near or far it will be i ... \\
        \cdashlinelr{1-3}
        & Reference & ... thou again meet any one after this hour reciting aught of poetry whether he be near or far it will be i ... \\
        \bottomrule
    \end{tabular}
    \vspace{-0.2cm}
    \caption{Example inference process of BECTRA (Alg.~\ref{algo:bectra_inference}), recognizing an utterance from LibriSpeech test-other set (2033-164914-0016). In BERT-CTC decoding (Alg.~\ref{algo:bectra_inference} line 2), the highlighted tokens are replaced with the mask token $\varnothing$ and repredicted in the next iteration. The BERT-CTC result is further refined via the transducer decoding (Alg.~\ref{algo:bectra_inference} line 3). Blue indicates corrected tokens, and red indicates ones not.}
    \vspace{-0.2cm}
    \label{tb:decoding_example}
\end{table*}

\paragraph{Overview}
Figure~\ref{fig:bectra} illustrates the proposed model,
namely \textbf{BE}RT-\textbf{C}TC-\textbf{Tra}nsducer (BECTRA),
that formulates E2E-ASR based on the transducer~\cite{graves2012sequence} and BERT-CTC.
BECTRA constructs an encoder as the BERT-CTC model,
where the output of the concatenation network in Eq.~\eqref{eq:p_bc_at} is used to calculate the transducer loss.
The joint and prediction networks are trained with an ASR-specific vocabulary, here a vocabulary constructed from the transcription of a target corpus.
BECTRA enables more suitable training of E2E-ASR compared to BERT-CTC,
unconstrained from the BERT vocabulary.

Let $W^{\mathsf{a}}\!=\!( w^{\mathsf{a}}_m\!\in\!\mathcal{V}^{\mathsf{a}} | m\!=\!1,\cdots,M)$ be an $M$-length output sequence corresponding to an input sequence $O$ and output sequence in the BERT vocabulary $W^{\mathsf{b}}$, where
$\mathcal{V}^{\mathsf{a}}$ is a vocabulary constructed from ASR training text.
The superscript ($\mathsf{a}$) indicates the use of the ASR vocabulary.
Similar to Eq.~\eqref{eq:p_bc_W_O},
BECTRA formulates E2E-ASR by marginalizing the conditional probability $p(W^{\mathsf{a}}|O)$ over all possible masked sequences as
\begin{equation}
    p(W^{\mathsf{a}}|O) = \sum_{\tilde{W}^{\mathsf{b}}} p(W^{\mathsf{a}}|\tilde{W}^{\mathsf{b}},O) p(\tilde{W}^{\mathsf{b}}|O).
    \vspace{-0.05cm}
    \label{eq:p_bt_W_O}
\end{equation}
Note that, in Eq.~\eqref{eq:p_bt_W_O}, $\tilde{W}^{\mathsf{b}}$ is obtained by masking a sequence in the BERT unit $W^{\mathsf{b}}$ ($\neq W^{\mathsf{a}}$).
As in Eq.~\eqref{eq:p_bc_W_tW_O},
$p(W^{\mathsf{a}}|\tilde{W}^{\mathsf{b}},O)$ is factorized
by considering possible alignment paths and making conditional independence assumptions as
\begin{equation}
    p(W^{\mathsf{a}}|\tilde{W}^{\mathsf{b}},O) \approx \sum_{A^{\mathsf{a}}\in\mathcal{B'}^{\text{--}1}(W^{\mathsf{a}})}
    p(A^{\mathsf{a}}|W^{\mathsf{a}},O) p(W^{\mathsf{a}}|\tilde{W}^{\mathsf{b}}),
    \label{eq:p_bt_W_tW_O}
\end{equation}
where $A^{\mathsf{a}}\!=\!(a^{\mathsf{a}}_u\!\in\!\mathcal{V}^{\mathsf{a}}\cup\{\epsilon\}|u\!=\!1,\cdots,T\!+\!M)$ is an alignment path between $O$ and $W^{\mathsf{a}}$,
defined by the transducer~\cite{graves2012sequence}, and
$\mathcal{B}':A^{\mathsf{a}}\!\mapsto\!W^{\mathsf{a}}$ is the collapsing function.
The alignment probability $p(A^{\mathsf{a}}|W^{\mathsf{a}},O)$ is further factorized by the probabilistic chain rule 
\textit{without a conditional independence assumption} (cf. Eq.~\eqref{eq:p_bc_W_tW_O_2}) as
\begin{align}
    &\hspace{-0.16cm}\resizebox{7.5cm}{!}{$\displaystyle
    \text{Eq.~\eqref{eq:p_bt_W_tW_O}} = \sum_{A^{\mathsf{a}}} \prod_{u=1}^{T+M} p(a^{\mathsf{a}}_u|a^{\mathsf{a}}_1,\cdots,a^{\mathsf{a}}_{u-1},W^{\mathsf{a}},O) p(W^{\mathsf{a}}|\tilde{W}^{\mathsf{b}}),$} \\
    &\hspace{-0.16cm}\resizebox{7.9cm}{!}{$\displaystyle
    \text{\phantom{Eq.~\eqref{eq:p_bt_W_tW_O}}} \approx \sum_{A^{\mathsf{a}}} \prod_{u=1}^{T+M} p(a^{\mathsf{a}}_u|\underbrace{w^{\mathsf{a}}_1,\cdots,w^{\mathsf{a}}_{m_{u}-1}}_{=\mathcal{B}'(a^{\mathsf{a}}_1,\cdots,a^{\mathsf{a}}_{u-1})},W^{\mathsf{a}},O) p(W^{\mathsf{a}}|\tilde{W}^{\mathsf{b}})$,}
    \label{eq:p_bt_W_tW_O_2b}
\end{align}
where $m_u$ is the number of tokens predicted up to an index of $u$.
In Eq.~\eqref{eq:p_bt_W_tW_O_2b},
we approximate $(a^{\mathsf{a}}_1,\cdots,a^{\mathsf{a}}_{u-1})\!\approx\!(w^{\mathsf{a}}_1,\cdots,w^{\mathsf{a}}_{m_u-1})$ as performed in the transducer,
which is reasonable since $W^{\mathsf{a}}$ can be decided uniquely by the collapsing function.
Similar to BERT-CTC in Eq.~\eqref{eq:p_bc_W_tW_O_3},
BECTRA models Eq.~\eqref{eq:p_bt_W_tW_O_2b} as
\begin{equation}
    \text{Eq.~\eqref{eq:p_bt_W_tW_O_2b}} \triangleq \sum_{A^{\mathsf{a}}} \prod_{u} p(a^{\mathsf{a}}_u|w^{\mathsf{a}}_{<m_{u}},\text{BERT}(\tilde{W}^{\mathsf{b}}),O), \label{eq:p_bt_W_tW_O_3}
\end{equation}
where we assume $p(W^{\mathsf{a}}|\tilde{W}^{\mathsf{b}})$ can be modeled by BERT,
as $W^{\mathsf{b}}$ and $W^{\mathsf{a}}$ are convertible to each other via their tokenizers.

The architecture of BECTRA is shown in Fig.~\ref{fig:bectra}, and
the token emission probability in Eq.~\eqref{eq:p_bt_W_tW_O_3} is computed as
\begin{align}
    &p(a^{\mathsf{a}}_u|w^{\mathsf{a}}_{<m_{u}},\text{BERT}(\tilde{W}^{\mathsf{b}}),O) \nonumber \\ &\hspace{0.5cm}=\sigma(\text{JointNet}(\text{ConcatNet}_t(E,H),\bm{\mathrm{q}}_{m_u})) \in [0,1]^{|\mathcal{V}^{\mathsf{a}}|+1}, \hspace{-0.1cm} \label{eq:p_bt_at} \\
    &\hspace{-0.056cm}\bm{\mathrm{q}}_{m_u}\!= \text{PredictionNet} (w^{\mathsf{a}}_1,\cdots,w^{\mathsf{a}}_{m_u-1}) \in \mathbb{R}^D, \label{eq:q_m}
\end{align}
where $\bm{\mathrm{q}}_{m_u}$ is a $D$-dimensional hidden vector
obtained by encoding the previous non-blank tokens using the prediction network.
In Eq.~\eqref{eq:p_bt_at},
the joint network maps the combined outputs of the concatenation and prediction networks into a joint space.
The adoption of the prediction network allows the model to explicitly capture causal dependency in output tokens,
which is one of the key differences 
BECTRA has an advantage over BERT-CTC.

\paragraph{Training}
By substituting Eq.~\eqref{eq:p_bt_W_tW_O_3} into Eq.~\eqref{eq:p_bt_W_O},
the transducer loss of BECTRA is derived in the same manner as Eq.~\eqref{eq:L_bec} as
\begin{equation}
    \mathcal{L}_{\mathsf{tra}} \triangleq -\mathbb{E}_{\tilde{W}^{\mathsf{b}} \sim \mathcal{M}(W^{\mathsf{b}})} \bigg[ \log \sum_{A^{\mathsf{a}}} \prod_{u} p(a^{\mathsf{a}}_u|w^{\mathsf{a}}_{<m_{u}},\text{BERT}(\tilde{W}^{\mathsf{b}}),O) \bigg], \label{eq:L_tra}
\end{equation}
where the summation over $A^{\mathsf{a}}$ is efficiently computed via dynamic programming~\cite{graves2012sequence}.
The objective function of BECTRA is defined by combining $\mathcal{L}_{\mathsf{bec}}$ from Eq.~\eqref{eq:L_bec} and $\mathcal{L}_{\mathsf{tra}}$ from Eq.~\eqref{eq:L_tra} as
\begin{equation}
    \mathcal{L}_{\mathsf{bectra}} = (1 - \lambda) \mathcal{L}_{\mathsf{bec}} + \lambda \mathcal{L}_{\mathsf{tra}},
    \label{eq:L_bt}
\end{equation}
where $\lambda\!\in\!(0,1)$ is a tunable parameter.

\begin{algorithm}[t]
    \caption{Decoding algorithm of BERT-CTC}
    \label{algo:bertctc_inference}
    \begin{algorithmic}[1]
        \algnotext{EndFor}
        \algnotext{EndFunction}
        \footnotesize
        \Function{DecodeBertctc}{$E$, $K$}
        \State Initilize a hypothesis $\hat{W}^{\mathsf{b}}$ with all mask tokens
        \For {$k=1,\dots,K$}
            \State Calculate $H=\text{BERT}(\hat{W}^{\mathsf{b}})$
            \State Update $\hat{W}^{\mathsf{b}}$ via the best path decoding based on Eq.~\eqref{eq:p_bc_at}
            \State Calculate the number of masked tokens as $M = \lfloor |\hat{W}^{\mathsf{b}}| \cdot \frac{K - k}{K} \rfloor$
            \State Update $\hat{W}^{\mathsf{b}}$ by masking $M$ tokens with the lowest probabilities
        \EndFor
        \State \Return $\hat{W}^{\mathsf{b}}$, $H$
        \EndFunction
    \end{algorithmic}
\end{algorithm}
\begin{algorithm}[t]
    \caption{Decoding algorithm of BECTRA}
    \label{algo:bectra_inference}
    \begin{algorithmic}[1]
        \algnotext{EndFunction}
        \footnotesize
        \Function{DecodeBectra}{$E$, $K$, $B$}
        \State Perform \Call{DecodeBertctc}{$E$, $K$} and obtain $H$
        \State Predict $\hat{W}^{\mathsf{a}}$ via the beam search decoding with a beam size of $B$,
        \Statex \qquad\quad using the joint and prediction networks in Eqs.~\eqref{eq:p_bt_at} and~\eqref{eq:q_m} 
        \State \Return $\hat{W}^{\mathsf{a}}$
        \EndFunction
    \end{algorithmic}
\end{algorithm}

\paragraph{Inference} Algorithm~\ref{algo:bectra_inference} shows the inference algorithm of BECTRA,
implemented with BERT-CTC decoding followed by beam-search decoding of the transducer~\cite{graves2012sequence,boyer2021study} (see Table~\ref{tb:decoding_example} for example).
BERT-CTC decoding provides the model with a fully contextualized BERT output $H$,
which is from the final hypothesis estimated by the iterative refinement (line 2).
The beam-search decoding uses the token emission probability from Eq.~\eqref{eq:p_bt_at} to find an optimal sequence with the highest sequence-level generation probability (line 3).
With this combined inference algorithm,
BECTRA can utilize BERT to capture bi-directional context in an output sequence,
the advantage of non-autoregressive decoding.
Moreover, the transducer-based decoding enables the model to further refine a sequence in an autoregressive manner,
using a more suitable output unit for ASR.

\begin{table*}[t]
    \centering
    \caption{Word or character error rates [\%] ($\downarrow$) of our proposed BECTRA compared to Conformer-Transducer (Cfm-T) and BERT-CTC baselines. Each model was trained using either a vocabulary used in BERT ($\mathcal{V}^{\mathsf{b}}$) or vocabulary constructed from ASR training text ($\mathcal{V}^{\mathsf{a}}$).}
    \label{tb:main_results}
    \resizebox{.88\linewidth}{!}{
    \begin{tabular}{lcccccccccx{0.8cm}x{0.8cm}x{0.8cm}x{0.8cm}}
        \toprule
        & & \multicolumn{4}{c}{\textbf{LibriSpeech-100h}} & \multicolumn{4}{c}{\textbf{LibriSpeech-960h}} & \multicolumn{2}{c}{\textbf{TED-LIUM2}} & \multicolumn{2}{c}{\textbf{AISHELL-1}} \\
        \cmidrule(l{0.3em}r{0.3em}){3-6}\cmidrule(l{0.3em}r{0.3em}){7-10}\cmidrule(l{0.3em}r{0.3em}){11-12}\cmidrule(l{0.3em}r{0.3em}){13-14}
        & \multirow{2}{*}[-5.8pt]{\shortstack[c]{\textbf{Output}\\\textbf{Vocab.}}}
        & \multicolumn{2}{c}{Dev WER} & \multicolumn{2}{c}{Test WER} & \multicolumn{2}{c}{Dev WER} & \multicolumn{2}{c}{Test WER} & \multirow{2}{*}[-3.2pt]{\shortstack[c]{Dev\\WER}} & \multirow{2}{*}[-3.2pt]{\shortstack[c]{Test\\WER}} & \multirow{2}{*}[-3.2pt]{\shortstack[c]{Dev\\CER}} & \multirow{2}{*}[-3.2pt]{\shortstack[c]{Test\\CER}} \\
        \cmidrule(l{0.3em}r{0.3em}){3-4}\cmidrule(l{0.3em}r{0.3em}){5-6}\cmidrule(l{0.3em}r{0.3em}){7-8}\cmidrule(l{0.3em}r{0.3em}){9-10}
        \textbf{Model} & & clean & other & clean & other & clean & other & clean & other \\
        \midrule
        Cfm-T & $\mathcal{V}^{\mathsf{b}}$ & 9.7 & 21.5 & 9.8 & 22.3 & -- & -- & -- & -- & -- & -- & -- & -- \\
        Cfm-T & $\mathcal{V}^{\mathsf{a}}$ & 5.9 & 17.7 & 6.0 & 17.6 & \textbf{2.5} & 6.8 & \textbf{2.8} & 6.8 & 7.8 & 7.4 & 4.9 & 5.3 \\
        BERT-CTC & $\mathcal{V}^{\mathsf{b}}$ & 7.0 & 16.4 & 7.1 & 16.5 & 3.1 & 7.1 & 3.2 & 7.1 & 8.3 & 7.6 & 3.9 & 4.0 \\
        BECTRA & $\mathcal{V}^{\mathsf{a}}$ & \textbf{5.1} & \textbf{15.4} & \textbf{5.4} & \textbf{15.5} & 2.6 & \textbf{6.7} & 2.9 & \textbf{6.7} & \textbf{7.3} & \textbf{6.9} & \textbf{3.7} & \textbf{3.9} \\
        \bottomrule
    \end{tabular}}
    \vspace{-0.2cm}
\end{table*}

\vspace{-0.25cm}
\section{Experiments}
\vspace{-0.2cm}

\subsection{Experimental Setting}
\vspace{-0.15cm}
We used the ESPnet toolkit~\cite{watanabe2018espnet} for conducting the experiments, and
all the codes and recipes are made publicly available at \url{https://github.com/YosukeHiguchi/espnet/tree/bectra}.

\paragraph{Data}
We evaluated models using various corpora with different amounts of data, speaking styles, and languages,
including
LibriSpeech (LS)~\cite{panayotov2015librispeech},
TED-LIUM2 (TED2)~\cite{rousseau2014enhancing},
and AISHELL-1 (AS1)~\cite{bu2017aishell}.
We also trained models on
the \textit{train-clean-100} subset of LS (LS-100)
for performing additional investigations and analyses.
To obtain the vocabulary $\mathcal{V}^{\mathsf{a}}$ from ASR transcriptions,
we used SentencePiece~\cite{kudo2018subword} to construct subword vocabularies for LS-100, LS and TED2,
where each vocabulary size $|\mathcal{V}^{\mathsf{a}}|$ was set to $300$, $5\text{k}$, and $500$, respectively.
For AS1,
we used character-level tokenization with $4231$ Chinese characters.

\paragraph{Evaluated models}
For a baseline model,
we trained Conformer-Transducer (Cfm-T) implemented in ESPnet~\cite{boyer2021study},
which is similar to the model in Fig.~\ref{fig:bectra} but without the concatenation, CTC, and BERT components.
BERT-CTC~\cite{higuchi2022bert} is also a baseline model trained based on $\mathcal{L}_{\mathsf{bec}}$ from Eq.~\eqref{eq:L_bec}.
BECTRA is the proposed model trained based on $\mathcal{L}_{\mathsf{bectra}}$ from Eq.~\eqref{eq:L_bt}.

\paragraph{Network architecture}
For the audio encoder,
we constructed the Conformer architecture~\cite{gulati2020conformer},
which consisted of two convolutional neural network layers followed by a stack of $12$ encoder blocks.
The number of heads $d_{\mathsf{h}}$, dimension of a self-attention layer $d_{\mathsf{model}}$, 
dimension of a feed-forward network $d_{\mathsf{ff}}$, and kernel size $K$ were set to $4$, $256$, $1024$, and $31$, respectively.
For the transducer-based models,
the prediction network was a single long short-term memory layer with $256$ hidden units.
For the models using BERT,
the concatenation network was the Transformer encoder~\cite{vaswani2017attention} with $6$ blocks, and
$d_{\mathsf{h}}$, $d_{\mathsf{model}}$, and $d_{\mathsf{ff}}$ were set to
$4$, $256$, and $2048$, respectively.
We used a BERT$_{\text{BASE}}$ of each language provided in HuggingFace~\cite{transformers2020wolf}:
English~\cite{bert_eng} ($|\mathcal{V}^{\mathsf{b}}|\!=\!30522$) and Mandarin~\cite{bert_man} ($|\mathcal{V}^{\mathsf{b}}|\!=\!21128$).
The output dimension $D$ of the audio encoder $\bm{\mathrm{e}}_t$, BERT $\bm{\mathrm{h}}_n$, and prediction network $\bm{\mathrm{q}}_{m_u}$ were all adjusted to match $d_{\mathsf{model}}$ ($=256$).

\paragraph{Training and decoding}
We mostly followed configurations provided by the ESPnet2 recipe for each dataset.
The models were trained to 100 epochs for LS-100 and AS1,
and 70 epochs for TED2 and LS.
The Adam optimizer~\cite{kingma2015adam} with Noam learning rate scheduling~\cite{vaswani2017attention} was used for weight updates,
where warmup steps and a peak learning rate were set to $15$k and $2\text{e-}3$, respectively.
We augmented speech data using speed perturbation~\cite{ko2015audio} with a factor of $3$ and SpecAugment~\cite{park2019specaugment}.
Similar to~\cite{lee2022memory},
we adopted the intermediate CTC regularization~\cite{lee2021intermediate} for the audio encoder,
where an auxiliary CTC loss was applied to the $6$-th layer and calculated based on the ASR vocabulary $\mathcal{V}^{\mathsf{a}}$.
For BECTRA training,
we set $\lambda$ to $0.5$ in Eq.~\eqref{eq:L_bt}.
After training, a final model was obtained for evaluation by averaging model parameters over ten checkpoints with the best validation performance.
For the number of iterations in BERT-CTC decoding (in Alg.~\ref{algo:bertctc_inference}),
we set $K$ to $20$ for BERT-CTC and $10$ for BECTRA.
We performed the beam search decoding with a beam size of $10$ for Cfm-T and $5$ for BECTRA.

\vspace{-0.27cm}
\subsection{Difficulty of Training ASR with BERT Vocabulary}
\vspace{-0.13cm}
\label{ssec:exp_bert_vocab}
In Table~\ref{tb:main_results},
we compare LS-100 results on Cfm-T trained with the 
the BERT or ASR vocabulary ($\mathcal{V}^{\mathsf{b}}$ vs.\ $\mathcal{V}^{\mathsf{a}}$).
At a glance, Cfm-T resulted in significantly worse WERs by using the BERT vocabulary,
suggesting unsuitable ASR training with the word-level and domain-mismatched BERT units.
We also mention that the large vocabulary size leads to increasingly high memory consumption during the transducer training~\cite{lee2022memory},
which makes it difficult to train a model, especially on large datasets.
In contrast,
BERT-CTC in Table~\ref{tb:main_results}
achieved decent results by effectively dealing with the vocabulary mismatch,
explicitly using BERT information~\cite{higuchi2022bert}.

\vspace{-0.27cm}
\subsection{Main Results}
\vspace{-0.13cm}
\label{ssec:exp_main}
Table~\ref{tb:main_results} lists the main results of all the tasks evaluated in the word error rate (WER)
or character error rate (CER).
Comparing Cfm-T (with $\mathcal{V}^{\mathsf{a}}$) and BERT-CTC,
BERT-CTC performed worse than Cfm-T in several tasks due to the vocabulary discrepancy.
For example, in TED2,
a severe domain mismatch exists between the BERT training text and ASR transcription (i.e., Wikipedia vs.\ lecture)
that caused unsuitable ASR training for BERT-CTC.
Overall,
BECTRA achieved the best result across all of the tasks.
The gain from Cfm-T demonstrates the effectiveness of exploiting BERT knowledge via the BERT-CTC-based encoder.
Moreover,
BECTRA improved BERT-CTC by generating text in a suitable output unit and
mitigating the conditional independence assumption (Eq.~\eqref{eq:p_bc_W_tW_O_2} vs.\ Eq.~\eqref{eq:p_bt_W_tW_O_2b}).
Another notable observation was that, with more data in LS-960,
the performance gap between Cfm-T and BECTRA was reduced, and
the usage of BERT became less influential.
We assume that LS-960 contained enough text data that
the models were able to learn rich linguistic information in the LibriSpeech-specific domain.

Table~\ref{tb:decoding_example} shows an example decoding process of BECTRA.
BERT-CTC succeeded at resolving most of the substitution errors via iterative refinement using BERT.
The transducer decoding was particularly effective at adjusting
deletion and insertion errors in BERT-CTC outputs, e.g., 
``poetryry''$\rightarrow$``poetry'' in Table~\ref{tb:decoding_example}.

\vspace{-0.1cm}
\subsection{Trade-off between WER and Inference Speed}
\vspace{-0.1cm}

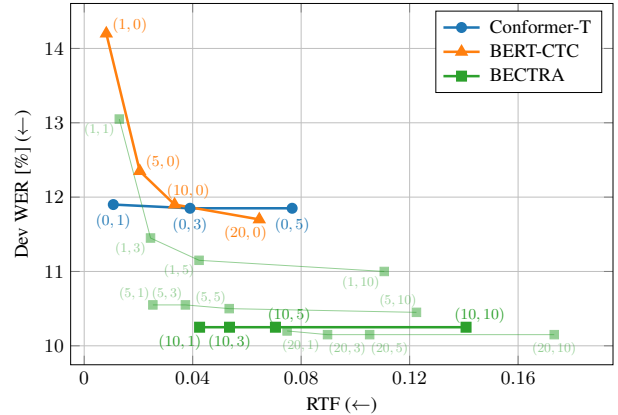
\begin{figure}[t]
\centering
\resizebox{0.95\columnwidth}{!}{
\begin{tikzpicture}
\definecolor{clr1}{RGB}{31, 119, 180}
\definecolor{clr2}{RGB}{255, 127, 14}
\definecolor{clr3}{RGB}{44, 160, 44}

\pgfplotsset{
  scale only axis,
}

\begin{axis}[
    height=6cm,
    width=9cm,
    grid=major,
    xlabel=RTF ($\leftarrow$),
    ylabel=Dev WER $\text{[\%]}$ ($\leftarrow$),
    xlabel style={font=\normalsize},
    ylabel style={font=\normalsize},
    xmin=-0.005,xmax=0.195,
    xtick={0,0.04,0.08,0.12,0.16},
    every tick label/.append style={font=\normalsize},
    legend cell align={left},
    scaled ticks=false,
    tick label style={/pgf/number format/fixed},
]

\addplot[
    mark=*, clr1,
    scatter/classes={a={clr1}},
    scatter src=explicit symbolic,
    nodes near coords*={\Label},
    every node near coord/.append style={scale=0.8,anchor=90,yshift=-2pt},
    visualization depends on={value \thisrow{label} \as \Label},
    very thick,
] table [meta=class] {rnnt.dat}; \label{plot_rnnt}

\addplot[
    mark=triangle*, clr2,
    mark size=2.5pt,
    scatter src=explicit symbolic,
    nodes near coords,
    every node near coord/.append style={scale=0.8,anchor=\Anc,yshift=0pt},
    visualization depends on={value \thisrow{anc} \as \Anc},
    very thick,
] table [meta=label] {bertctc.dat}; \label{plot_bertctc}

\addplot[
    mark=square*, clr3,
    scatter src=explicit symbolic,
    nodes near coords,
    every node near coord/.append style={scale=0.8,anchor=\Anc,yshift=-1pt},
    visualization depends on={value \thisrow{anc} \as \Anc},
    very thick,
] table [meta=label] {bectra10.dat}; \label{plot_bectra}

\addplot[
    mark=square*, clr3,
    opacity=0.5,
    scatter/classes={c={clr3}},
    scatter src=explicit symbolic,
    nodes near coords*={\Label},
    every node near coord/.append style={scale=0.7,anchor=20,yshift=-2pt},
    visualization depends on={value \thisrow{label} \as \Label},
    thin,
] table [meta=class] {bectra1.dat};

\addplot[
    mark=square*, clr3,
    opacity=0.5,
    scatter/classes={c={clr3}},
    scatter src=explicit symbolic,
    nodes near coords*={\Label},
    every node near coord/.append style={scale=0.7,anchor=300,xshift=-8pt,yshift=0pt},
    visualization depends on={value \thisrow{label} \as \Label},
    thin,
] table [meta=class] {bectra5.dat};

\addplot[
    mark=square*, clr3,
    opacity=0.5,
    scatter src=explicit symbolic,
    nodes near coords,
    every node near coord/.append style={scale=0.7,anchor=\Anc,yshift=-2pt},
    visualization depends on={value \thisrow{anc} \as \Anc},
    thin,
] table [meta=label] {bectra20.dat};

\addlegendimage{/pgfplots/refstyle=plot_rnnt}\addlegendentry{\ Conformer-T}
\addlegendimage{/pgfplots/refstyle=plot_bertctc}\addlegendentry{\ BERT-CTC}
\addlegendimage{/pgfplots/refstyle=plot_bectra}\addlegendentry{\ BECTRA}

\end{axis}

\end{tikzpicture}
}
\vspace{-0.35cm}
\caption{Trade-off between WER and RTF on LS dev.\ sets. The values in parenthesis show the number of iterations and a beam size $(K, B)$.}
\vspace{-0.05cm}

\label{fig:iterations}
\end{figure}

Figure~\ref{fig:iterations} depicts the trade-off between WER and real-time factor (RTF) on the LS dev-\{clean, other\} sets.
RTF was measured using a single V100 GPU with a batchsize of $1$.
Looking at the results from greedy decoding ($K\!\le\!1$ and $B\!=\!1$),
Cfm-T resulted in the best performance with the lowest WER and RTF.
The performance gain was more substantial for increasing the number of iterations in BERT-CTC
than for increasing a beam size in Cfm-T.
BECTRA greatly benefited from beam search decoding at $K\!=\!1$, but the improvement became small with more iterations $K\!>\!1$.
This indicates that BERT-CTC decoding refined the output sequence
enough that the search space was reduced during beam search decoding, and
only a small beam size ($K\!=\!3$ or $5$) is adequate without degrading the inference speed.
All in all, BECTRA with $K\!=\!10$ and $B\!\in\![1,5]$ resulted in the preferable performance balancing the trade-off reasonably well,
taking advantage of both fast non-autoregressive decoding of BERT-CTC and accurate autoregressive decoding of Cfm-T.

\vspace{-0.1cm}
\section{Conclusion}
\vspace{-0.1cm}
We proposed BECTRA, a novel E2E-ASR model formulated by the transducer and BERT-CTC.
BECTRA adopts BERT-CTC for the audio encoder,
aiming to exploit BERT for capturing contextual information.
In addition, prediction and joint networks are trained with an ASR-specific vocabulary,
which enables the model to perform ASR in a suitable unit.
One of the limitations of BECTRA is its non-streaming property due to the BERT-CTC's non-autoregressive formulation.
The possible solutions include the adoption of the two-pass modeling~\cite{sainath2019two} or blockwise-attention mechanism~\cite{wang2021streaming}.

\paragraph{Acknowledgement} This work was supported in part by JST ACT-X (JPMJAX210J) and JSPS KAKENHI (JP21J23495).

\newpage

\fontsize{8.7pt}{9.6pt}\selectfont
\bibliographystyle{IEEEtran}
\bibliography{refs}

\begin{thebibliography}{10}
\providecommand{\url}[1]{#1}
\csname url@samestyle\endcsname
\providecommand{\newblock}{\relax}
\providecommand{\bibinfo}[2]{#2}
\providecommand{\BIBentrySTDinterwordspacing}{\spaceskip=0pt\relax}
\providecommand{\BIBentryALTinterwordstretchfactor}{4}
\providecommand{\BIBentryALTinterwordspacing}{\spaceskip=\fontdimen2\font plus
\BIBentryALTinterwordstretchfactor\fontdimen3\font minus
  \fontdimen4\font\relax}
\providecommand{\BIBforeignlanguage}[2]{{%
\expandafter\ifx\csname l@#1\endcsname\relax
\typeout{** WARNING: IEEEtran.bst: No hyphenation pattern has been}%
\typeout{** loaded for the language `#1'. Using the pattern for}%
\typeout{** the default language instead.}%
\else
\language=\csname l@#1\endcsname
\fi
#2}}
\providecommand{\BIBdecl}{\relax}
\BIBdecl

\bibitem{devlin2019bert}
J.~Devlin, M.-W. Chang, K.~Lee, and K.~Toutanova, ``{BERT}: Pre-training of
  deep bidirectional {Transformers} for language understanding,'' in
  \emph{Proc. NAACL-HLT}, 2019, pp. 4171--4186.

\bibitem{brown2020language}
T.~Brown, B.~Mann, N.~Ryder, M.~Subbiah \emph{et~al.}, ``Language models are
  few-shot learners,'' in \emph{Proc. NeurIPS}, 2020, pp. 1877--1901.

\bibitem{tenney2019bert}
I.~Tenney, D.~Das, and E.~Pavlick, ``{BERT} rediscovers the classical {NLP}
  pipeline,'' in \emph{Proc. ACL}, 2019, pp. 4593--4601.

\bibitem{shin2019effective}
J.~Shin, Y.~Lee, and K.~Jung, ``Effective sentence scoring method using {BERT}
  for speech recognition,'' in \emph{Proc. ACML}, 2019, pp. 1081--1093.

\bibitem{salazar2020masked}
J.~Salazar, D.~Liang, T.~Q. Nguyen, and K.~Kirchhoff, ``Masked language model
  scoring,'' in \emph{Proc. ACL}, 2020, pp. 2699--2712.

\bibitem{chiu2021innovative}
S.-H. Chiu and B.~Chen, ``Innovative {BERT}-based reranking language models for
  speech recognition,'' in \emph{Proc. SLT}, 2021, pp. 266--271.

\bibitem{futami2021asr}
H.~Futami, H.~Inaguma, M.~Mimura, S.~Sakai \emph{et~al.}, ``{ASR} rescoring and
  confidence estimation with {ELECTRA},'' in \emph{Proc. ASRU}, 2021, pp.
  380--387.

\bibitem{udagawa2022effect}
T.~Udagawa, M.~Suzuki, G.~Kurata, N.~Itoh \emph{et~al.}, ``Effect and analysis
  of large-scale language model rescoring on competitive {ASR} systems,'' in
  \emph{Proc. Interspeech}, 2022, pp. 3919--3923.

\bibitem{futami2020distilling}
H.~Futami, H.~Inaguma, S.~Ueno, M.~Mimura \emph{et~al.}, ``Distilling the
  knowledge of {BERT} for sequence-to-sequence {ASR},'' in \emph{Proc.
  Interspeech}, 2020, pp. 3635--3639.

\bibitem{bai2021fast}
Y.~Bai, J.~Yi, J.~Tao, Z.~Tian \emph{et~al.}, ``Fast end-to-end speech
  recognition via non-autoregressive models and cross-modal knowledge
  transferring from {BERT},'' \emph{IEEE/ACM Trans. Audio, Speech, Lang.
  Process.}, vol.~29, pp. 1897--1911, 2021.

\bibitem{kubo2022knowledge}
Y.~Kubo, S.~Karita, and M.~Bacchiani, ``Knowledge transfer from large-scale
  pretrained language models to end-to-end speech recognizers,'' in \emph{Proc.
  ICASSP}, 2022, pp. 8512--8516.

\bibitem{lu2022context}
K.-H. Lu and K.-Y. Chen, ``A context-aware knowledge transferring strategy for
  {CTC}-based {ASR},'' in \emph{Proc. SLT}, 2022.

\bibitem{huang2021speech}
W.-C. Huang, C.-H. Wu, S.-B. Luo, K.-Y. Chen \emph{et~al.}, ``Speech
  recognition by simply fine-tuning {BERT},'' in \emph{Proc. ICASSP}, 2021, pp.
  7343--7347.

\bibitem{yi2021efficiently}
C.~Yi, S.~Zhou, and B.~Xu, ``Efficiently fusing pretrained acoustic and
  linguistic encoders for low-resource speech recognition,'' \emph{IEEE Signal
  Process. Lett.}, vol.~28, pp. 788--792, 2021.

\bibitem{zheng2021wav}
G.~Zheng, Y.~Xiao, K.~Gong, P.~Zhou \emph{et~al.}, ``Wav-{BERT}: Cooperative
  acoustic and linguistic representation learning for low-resource speech
  recognition,'' in \emph{Proc. Findings of EMNLP}, 2021, pp. 2765--2777.

\bibitem{deng2021improving}
K.~Deng, S.~Cao, Y.~Zhang, and L.~Ma, ``Improving hybrid {CTC}/attention
  end-to-end speech recognition with pretrained acoustic and language models,''
  in \emph{Proc. ASRU}, 2021, pp. 76--82.

\bibitem{yu2022non}
F.-H. Yu, K.-Y. Chen, and K.-H. Lu, ``Non-autoregressive {ASR} modeling using
  pre-trained language models for {Chinese} speech recognition,''
  \emph{IEEE/ACM Trans. Audio, Speech, Lang. Process.}, vol.~30, pp.
  1474--1482, 2022.

\bibitem{higuchi2022bert}
Y.~Higuchi, B.~Yan, S.~Arora, T.~Ogawa \emph{et~al.}, ``{BERT} meets {CTC}: New
  formulation of end-to-end speech recognition with pre-trained masked language
  model,'' in \emph{Proc. Findings of EMNLP}, 2022.

\bibitem{ghazvininejad2019mask}
M.~Ghazvininejad, O.~Levy, Y.~Liu, and L.~Zettlemoyer, ``Mask-predict:
  {Parallel} decoding of conditional masked language models,'' in \emph{Proc.
  EMNLP-IJCNLP}, 2019, pp. 6114--6123.

\bibitem{deng2022improving}
K.~Deng, Z.~Yang, S.~Watanabe, Y.~Higuchi \emph{et~al.}, ``Improving
  non-autoregressive end-to-end speech recognition with pre-trained acoustic
  and language models,'' in \emph{Proc. ICASSP}, 2022, pp. 8522--8526.

\bibitem{graves2012sequence}
A.~Graves, ``Sequence transduction with recurrent neural networks,''
  \emph{arXiv preprint arXiv:1211.3711}, 2012.

\bibitem{graves2006connectionist}
A.~Graves, S.~Fern{\'a}ndez, F.~Gomez, and J.~Schmidhuber, ``Connectionist
  temporal classification: {L}abelling unsegmented sequence data with recurrent
  neural networks,'' in \emph{Proc. ICML}, 2006, pp. 369--376.

\bibitem{vaswani2017attention}
A.~Vaswani, N.~Shazeer, N.~Parmar, J.~Uszkoreit \emph{et~al.}, ``Attention is
  all you need,'' in \emph{Proc. NeurIPS}, 2017, pp. 5998--6008.

\bibitem{chan2020imputer}
W.~Chan, C.~Saharia, G.~Hinton, M.~Norouzi \emph{et~al.}, ``Imputer: Sequence
  modelling via imputation and dynamic programming,'' in \emph{Proc. ICML},
  2020, pp. 1403--1413.

\bibitem{higuchi2020mask}
Y.~Higuchi, S.~Watanabe, N.~Chen, T.~Ogawa \emph{et~al.}, ``{Mask CTC}:
  Non-autoregressive end-to-end {ASR} with {CTC} and mask predict,'' in
  \emph{Proc. Interspeech}, 2020, pp. 3655--3659.

\bibitem{higuchi2021improved}
Y.~Higuchi, H.~Inaguma, S.~Watanabe, T.~Ogawa \emph{et~al.}, ``Improved
  mask-{CTC} for non-autoregressive end-to-end {ASR},'' in \emph{Proc. ICASSP},
  2021, pp. 8363--8367.

\bibitem{boyer2021study}
F.~Boyer, Y.~Shinohara, T.~Ishii, H.~Inaguma \emph{et~al.}, ``A study of
  {Transducer} based end-to-end {ASR} with {ESPnet}: Architecture, auxiliary
  loss and decoding strategies,'' in \emph{Proc. ASRU}, 2021, pp. 16--23.

\bibitem{watanabe2018espnet}
S.~Watanabe, T.~Hori, S.~Karita, T.~Hayashi \emph{et~al.}, ``{ESPnet}:
  {E}nd-to-end speech processing toolkit,'' in \emph{proc. Interspeech}, 2018,
  pp. 2207--2211.

\bibitem{panayotov2015librispeech}
V.~Panayotov, G.~Chen, D.~Povey, and S.~Khudanpur, ``Librispeech: An {ASR}
  corpus based on public domain audio books,'' in \emph{Proc. ICASSP}, 2015,
  pp. 5206--5210.

\bibitem{rousseau2014enhancing}
A.~Rousseau, P.~Del{\'e}glise, and Y.~Est{\`e}ve, ``Enhancing the {TED}-{LIUM}
  corpus with selected data for language modeling and more {TED} talks,'' in
  \emph{Proc. LREC}, 2014, pp. 3935--3939.

\bibitem{bu2017aishell}
H.~Bu, J.~Du, X.~Na, B.~Wu \emph{et~al.}, ``{AISHELL}-1: An open-source
  {Mandarin} speech corpus and a speech recognition baseline,'' in \emph{Proc.
  O-COCOSDA}, 2017, pp. 1--5.

\bibitem{kudo2018subword}
T.~Kudo, ``Subword regularization: {I}mproving neural network translation
  models with multiple subword candidates,'' in \emph{Proc. ACL}, 2018, pp.
  66--75.

\bibitem{gulati2020conformer}
A.~Gulati, J.~Qin, C.-C. Chiu, N.~Parmar \emph{et~al.}, ``Conformer:
  {C}onvolution-augmented {Transformer} for speech recognition,'' in
  \emph{Proc. Interspeech}, 2020, pp. 5036--5040.

\bibitem{transformers2020wolf}
T.~Wolf, L.~Debut, V.~Sanh, J.~Chaumond \emph{et~al.}, ``Transformers:
  State-of-the-art natural language processing,'' in \emph{Proc. EMNLP: System
  Demonstrations}, 2020, pp. 38--45.

\bibitem{bert_eng}
``bert-base-uncased,'' \url{https://huggingface.co/bert-base-uncased}, [Online;
  Accessed on October-10-2022].

\bibitem{bert_man}
``bert-base-chinese,'' \url{https://huggingface.co/bert-base-chinese}, [Online;
  Accessed on October-10-2022].

\bibitem{kingma2015adam}
D.~P. Kingma and J.~Ba, ``Adam: A method for stochastic optimization,'' in
  \emph{Proc. ICLR}, 2015.

\bibitem{ko2015audio}
T.~Ko, V.~Peddinti, D.~Povey, and S.~Khudanpur, ``Audio augmentation for speech
  recognition,'' in \emph{Proc. Interspeech}, 2015, pp. 3586--3589.

\bibitem{park2019specaugment}
D.~S. Park, W.~Chan, Y.~Zhang, C.-C. Chiu \emph{et~al.}, ``{SpecAugment}: {A}
  simple data augmentation method for automatic speech recognition,'' in
  \emph{Proc. Interspeech}, 2019, pp. 2613--2617.

\bibitem{lee2022memory}
J.~Lee, L.~Lee, and S.~Watanabe, ``Memory-efficient training of
  {RNN-Transducer} with sampled softmax,'' in \emph{Proc. Interspeech}, 2022.

\bibitem{lee2021intermediate}
J.~Lee and S.~Watanabe, ``Intermediate loss regularization for {CTC}-based
  speech recognition,'' in \emph{Proc. ICASSP}, 2021, pp. 6224--6228.

\bibitem{sainath2019two}
T.~N. Sainath, R.~Pang, D.~Rybach, Y.~He \emph{et~al.}, ``Two-pass end-to-end
  speech recognition,'' in \emph{Proc. Interspeech}, 2019, pp. 2773--2777.

\bibitem{wang2021streaming}
T.~Wang, Y.~Fujita, X.~Chang, and S.~Watanabe, ``Streaming end-to-end {ASR}
  based on blockwise non-autoregressive models,'' in \emph{Proc. Interspeech},
  2021, pp. 3755--3759.

\end{thebibliography}

\end{document}